# Title: Rethinking Annotation Granularity for Overcoming Shortcuts in Deep Learning-based Radiograph Diagnosis: A Multicenter Study

Luyang Luo, PhD,[1] Hao Chen, PhD,[2] Yongjie Xiao, ME,[3] Yanning Zhou, PhD,[1] Xi Wang, PhD,[1] Varut Vardhanabhuti, PhD,[4] Mingxiang Wu, MM,[5] Chu Han, PhD,[6] Zaiyi Liu, MD,[6] Xin Hao Benjamin Fang, MD,[7] Efstratios Tsougenis, PhD,[8] Huangjing Lin, PhD,[1,3] Pheng-Ann Heng, PhD[1,9]

[1] Department of Computer Science and Engineering, The Chinese University of Hong Kong, Hong Kong, China.
[2] Department of Computer Science and Engineering, The Hong Kong University of Science and Technology, Hong Kong, China.
[3] AI Research Lab, Imsight Technology, Co., Ltd., Shenzhen, China.
[4] Department of Diagnostic Radiology, Li Ka Shing Faculty of Medicine, The University of Hong Kong, Hong Kong, China.
[5] Department of Radiology, Shenzhen People's Hospital, Luohu, Shenzhen, China.
[6] Department of Radiology, Guangdong Provincial People's Hospital, Guangdong Academy of Medical Sciences, Guangzhou, China.
[7] Department of Radiology, Queen Marry Hospital, Hong Kong, China.
[8] Artificial Intelligence Lab, Head Office Information Technology and Health Informatics Division, Hospital Authority, Hong Kong, China.
[9] Guangdong-Hong Kong-Macao Joint Laboratory of Human-Machine Intelligence-Synergy Systems, Shenzhen Institutes of Advanced Technology, Chinese Academy of Sciences, China.

**Correspondence Author:**

Dr. Hao Chen, Department of Computer Science and Engineering, 3/F Academic Building, The Hong Kong University of Science and Technology, Clear Water Bay, Kowloon, Hong Kong, China. Tel: +852-23588346. E-mail: jhc@cse.ust.hk.

**Funding Information**

This work was supported by Key-Area Research and Development Program of Guangdong Province, China (2020B010165004, 2018B010109006), Hong Kong Innovation and Technology Fund (Project No. ITS/311/18FP), HKUST Bridge Gap Fund (BGF.005.2021), National Natural Science Foundation of China with Project No. U1813204, and Shenzhen-HK Collaborative Development Zone.

**Article Type:** Original Research

**Summary Statement:** Fine-grained annotations (i.e., bounding boxes forsions) help chest radiograph diagnosis models overcome learning shortcuts by enabling the models to identify the correct lesion areas, leading to significantly improved radiograph-level classification performance.

**Key Points**

■ A deep learning model trained with chest radiograph-level annotations (CheXNet) achieved radiologist-level performance on the internal testing set, such as achieving an area under the receiver operating characteristic (ROC) curve [AUC] of 0.93 in classifying fracture, but made decisions from regions other than the true signs of the diseases, leading to dramatically degraded external performance for external test.

■ A deep learning model trained with fine-grained lesion-level annotations (CheXDet) also achieved radiologist-level performance on the internal testing set with significant improvement for external performance, such as in classifying fractures on the NIH-Google dataset (CheXDet AUC: 0.67 CheXNet AUC: 0.51; p<.001) and on the PadChest dataset (CheXDet AUC: 0.78, CheXNet AUC: 0.55; p<.01).

■ CheXDet achieved higher lesion localization performance than CheXNet for most abnormalities on all datasets, such as in detecting pneumothorax on the internal testing sets (CheXDet jacknife alternative free-respnse ROC-figure of merit [JAFROC-FOM]: 0.87, CheXNet JAFROC-FOM: 0.13; p<.001) and the external NIH-ChestX-ray14 dataset (CheXDet JAFROC-FOM: 0.55, CheXNet JAFROC-FOM: 0.04; p<.001).

**Abbreviations**
AUC = area under the receiver operating characteristic (ROC) curve
JAFROC-FOM = jacknife alternative free-respnse ROC-figure of merit
DL = Deep learning
DS1 = dataset 1
Grad-CAM = gradient-weighted class activation map

**Abstract**

**Purpose**

To evaluate the ability of fine-grained annotations to overcome shortcut learning in deep learning (DL)-based diagnosis using chest radiographs.

**Materials and Methods**

Two DL models were developed using radiograph-level annotations (yes or no disease) and fine-grained lesion-level annotations (lesion bounding boxes), respectively named CheXNet and CheXDet. A total of 34,501 chest radiographs from January 2005-September 2019 were retrospectively collected and annotated with cardiomegaly, pleural effusion, mass, nodule, pneumonia, pneumothorax, tuberculosis, fracture, and aortic calcification. The models' internal classification performance and lesion localization performance were compared on an testing set (n=2,922), external classification performance was compared on NIH-Google (n=4,376) and PadChest (n=24,536) datasets, and external lesion localization performance was compared on NIH-ChestX-ray14 dataset (n=880). The models were also compared to radiologists on a subset of the internal testing set (n=496). Performance was evaluated using receiver operating characteristic curve (ROC) analysis.

**Results**

Given sufficient training data, both models performed comparably to radiologists. CheXDet achieved significant improvement for external classification, such as in classifying fracture on NIH-Google (CheXDet area under the ROC curve [AUC]: 0.67, CheXNet AUC: 0.51; p<.001) and PadChest (CheXDet AUC: 0.78, CheXNet AUC: 0.55; p<.001). CheXDet achieved higher lesion detection performance than CheXNet for most abnormalities on all datasets, such as in detecting pneumothorax on the internal set (CheXDet jacknife alternative free-response ROC-figure of merit [JAFROC-FOM]: 0.87, CheXNet JAFROC-FOM: 0.13; p<.001) and NIH-ChestX-ray14 (CheXDet JAFROC-FOM: 0.55, CheXNet JAFROC-FOM: 0.04; p<.001).

**Conclusion**

Fine-grained annotations overcame shortcut learning and enabled DL models to identify correct lesion patterns, improving the models' generalizability.

## Introduction

As a driving force of the current technological transformation, robust and trustworthy artificial intelligence (AI) is in greater need than ever. Despite achieving expert-level accuracies on many diseases screening tasks (1-9), Deep Learning (DL)-based (10) AI models have been shown to make even correct decisions for the wrong reasons (11-13) and demonstrate considerably degraded performance when applied to external data (13-15). This phenomenon is referred to as shortcut learning (16), where deep neural networks unintendedly learned dataset biases (17) to fit the training data quickly. Specifically, dataset biases are the patterns that frequently co-occurred with the target disease and are more easily recognized than the true disease signs (18). While widely adopted DL diagnosis models are often developed with image-level binary annotations with "1" indicating the presence and "0" indicating the absence of disease, such spurious correlations could be captured by the DL model to fit the training data quickly (11, 19). For example, previous studies have found that DL-based classification models could rely on hospital tokens (see examples in Figure E1) to decide whether a chest radiograph contains pneumonia, fracture, or even COVID-19 lesions (12, 13, 20), leading to concerns about the DL models' credibility.

A possible way to alleviate shortcut learning is enlarging the model's learned distribution by incorporating more training data (15, 20). Previous works have also proposed using annotations such as bounding boxes of objects to constrain the DL models to learn from targeted regions (12, 21, 22). However, several questions are yet to be explored: would increasing training data always lead to a better disease classification model? Could fine-grained annotations alleviate shortcut learning and bring significant improvement to the DL models? More importantly, does overcoming shortcut learning help improve the DL models' generalizability on multicenter data?

In this study, we developed a classification model using radiograph-level annotations, CheXNet (4), and a detection model using lesion-level annotations, CheXDet, for an extensive comparison on the tasks of disease classification and lesion detection. We aimed to investigate the ability of fine-grained annotations on chest radiographs to improve DL model-based diagnosis.

## Materials and Methods

This retrospective study was approved by the institutional ethical committee (Approval No. YB-2021-554). The requirement for individual patient consent was waived, and all data from the institution were deidentified. Other data used for additional training or testing were publicly available. Figure 1 illustrates the construction and splitting of all datasets. This study followed the Standards for Reporting of Diagnostic Accuracy reporting guideline.

### Construction of Internal Dataset

We retrospectively collected 34,501 frontal-view chest radiographs and corresponding text reports of 30,561 patients from the clinical picture archiving and communication system from January 1, 2005 to September 31, 2019. This dataset is referred to as dataset 1 (DS1), where each radiograph was labeled yes or no for nine diseases (cardiomegaly, pleural effusion, mass, nodule, pneumonia, pneumothorax, tuberculosis, fracture, and aortic calcification) as well as containing the bounding boxes (i.e., fine-grained annotations) of the corresponding lesions. The radiographs were split into three different sets for training (n=28,673), tuning (n=2,906), and internal testing (n=2,922) without overlapping of patients. To assess the influence of the training data scale, we developed several different versions of the models using random samples containing 20% (n=5,763), 40% (n=11,493), 60% (n=17,180), 80% (n=22,943), and 100% of the training set. Also, a subset (n=496) was randomly sampled from the internal testing set to compare the performance between the models and radiologists.

### Groundtruth Labeling of Datset 1

For each radiograph in DS1, two readers were assigned for groundtruth labeling from a cohort of ten radiologists (4-30 years in general radiology). The chest radiographs and text reports were provided to the readers to label mentioned pathological findings and bounding boxes of the lesions. The radiologists' consensus with the text report was considered as the ground truth. For annotations of lesion bounding boxes, disagreements between the 2 readers were reviewed by another senior radiologist (at least 20 year-experience) from the cohort who made the final decision. These readers were not further involved in evaluation of model performance. All readers were provided with a graphical user interface-based annotation infrastructure. All images were kept the same size as their original digital imaging and communications in medicine (DICOM) format. The readers could zoom in and out using the software and change the window settings of the images, and images were viewed using monitors with resolutions equivalent to those used in clinical reporting. All readers were provided with the same guidelines for the annotation software and rules.

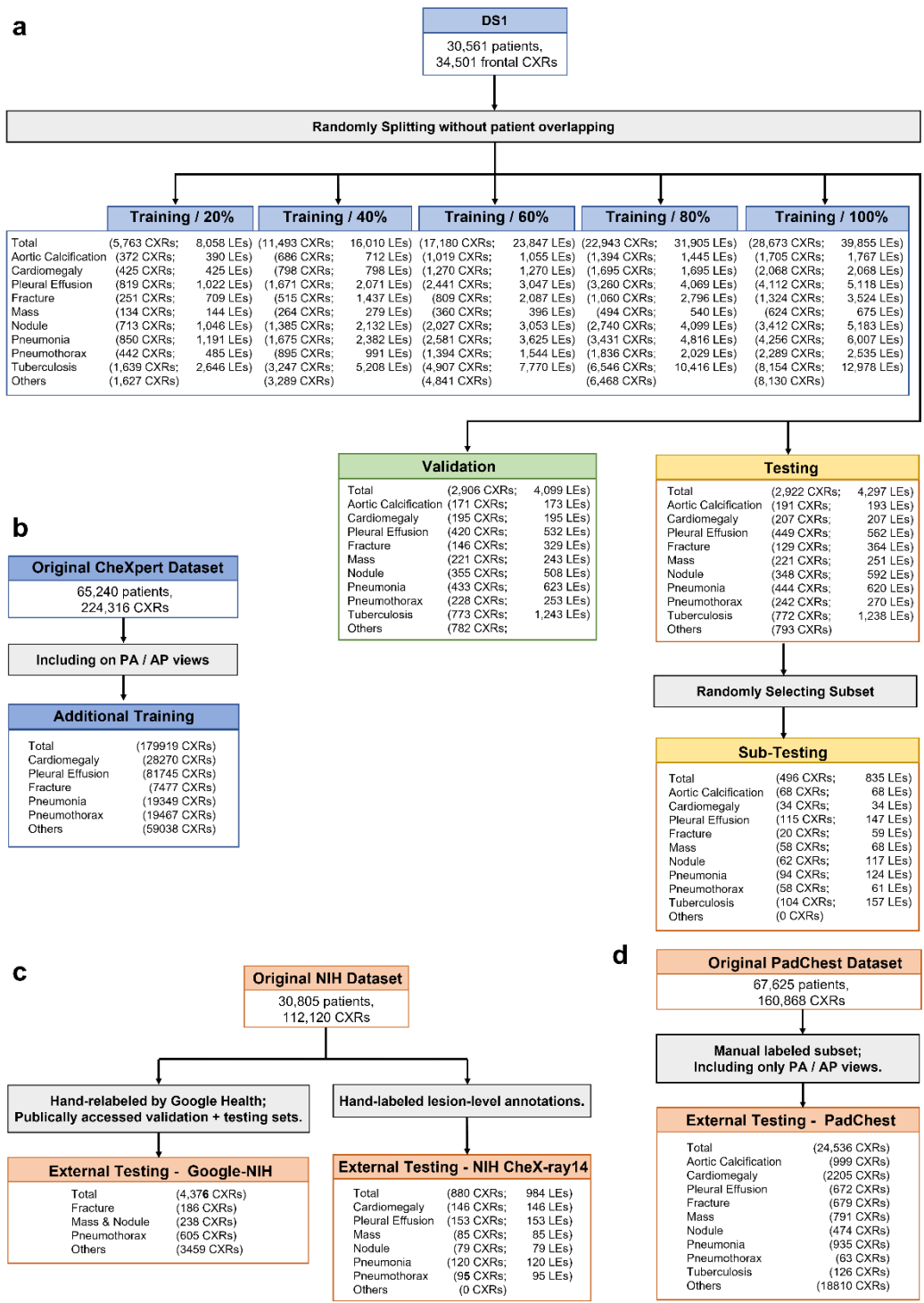

**Figure 1: Flowchart of images used from different cohorts.** (a) Split of the datset 1 (DS1), where five training sets containing different numbers of images and lesions were used for developing different versions of the models, a tuning set was used to select the best models, and a testing set was used for final evaluation. A subset was further randomly selected from the testing set for comparing the deep learning models with radiologists. (b) Frontal chest radiographs from the original CheXpert dataset were used as the additional training data. (c) Two subsets from the original NIH dataset were used for external testing. (d) The manually-labeled posterior-anterior/anterior-posterior (PA/AP) views from the PadChest dataset were used for external testing. CXRs = chest X-rays, LEs = lesions

*External Testing Datasets and Additional Training Data*

Three public-available datasets were used for external testing: 1) NIH-Google: A subset from the original National Institutes of Health (NIH) ChestX-ray14 database (23) containing a total of 4,376 frontal chest radiographs. Each radiograph was labeled with yes or no findings of airspace opacity, fracture, mass or nodule, or pneumothorax by at least three radiologists from Google Health (7). The latter three classes overlapped with those in DS1. 2) PadChest: A subset from the original PadChest (24) was used, containing 24,536 frontal radiographs labeled by trained physicians at the radiograph level. PadChest contained all nine classes in DS1. 3) NIH ChestX-ray14 dataset: 880 frontal chest radiographs with bounding box annotations of lesions hand-labeled by a board-certified radiologist were used, where six diseases (cardiomegaly, pleural effusion, nodule, mass, pneumonia, and pneumothorax) overlapped with the DS1's annotations.

Moreover, to evaluate whether increasing training data led to better performance for CheXNet (4), 179,919 frontal chest radiographs from the CheXpert dataset (25) were included as additional training data. This dataset was automatically annotated at the radiograph level with text reports by a natural language processing algorithm.

*Implementation of CheXNet and CheXDet*

CheXNet is a 121-layer densely-connected network (DenseNet-121) (26), which was trained with radiograph-level annotations to predict existence of the nine diseases (Figure E2). CheXNet contained four dense blocks, where the features from every shallow layer were concatenated and fed into the deeper layers for better gradient back-propagation. A convolutional layer and a pooling layer were appended after each dense block to conduct dimension reduction. The original output layer (1000-way softmax) of DenseNet-121 was replaced with a nine-way sigmoid layer (nine neurons, each of which was tailed with a sigmoid function to output disease probability).

CheXDet is a two-stage object detection network trained with lesion-level annotations to output lesion bounding boxes and the disease probabilities of suspected abnormal regions(Figure 2). In brief, CheXDet used EfficientNet (27) as the feature extractor and three bidirectional feature pyramid network (BiFPN) (28) layers for feature aggregation and enrichment. The BiFPN features were further fed into a region proposal network (RPN) (29) module and a region of interest (ROI) alignment module (30) for object proposal generation. The proposal features were further fed into four convolutional layers, and two fully connected layers were then used to conduct classification and bounding box regression based on the proposals, respectively.

Five versions of CheXNet and CheXDet were developed with 20%, 40%, 60%, 80%, and 100% of DS1 training data, respectively. For simplicity, we use subscripts in model names to indicate how much data were used to develop the model (e.g., CheXDet$_{20}$ indicates the CheXDet model developed with 20% of the training data). Moreover, we developed another version of CheXNet with training data from both DS1 and CheXpert, and this model is indicated as CheXNet$_{100+}$. All hyperparameters of the models were tuned on the tuning set (please find the details in sections 2.1 and 3.1 of the appendix). Figure 3 illustrates the brief training and testing processes of CheXNet and CheXDet. More details of the development processes for the two models can be found in appendix (sections 2 and 3).

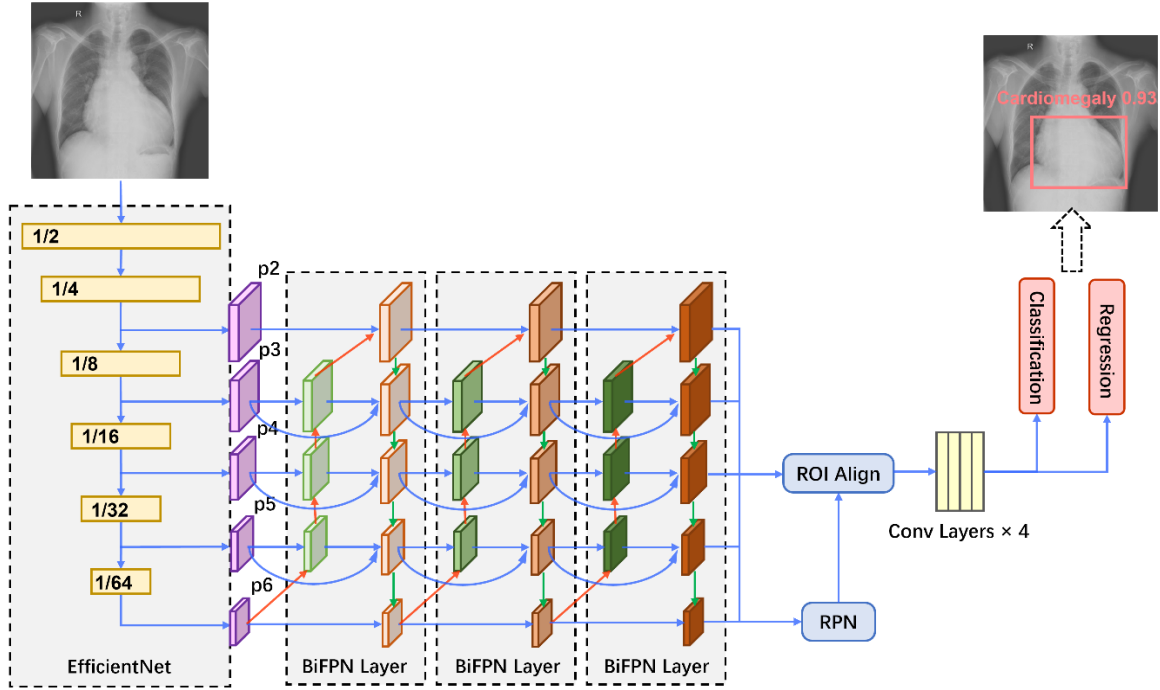

**Figure 2: CheXDet architecture.** An EfficientNet backbone is used for feature extraction, which also downsamples the data in width and height. The multiscale features (i.e., p2, p3, p4, p5, and p6) are then fed into three bi-directional feature pyramid network (BiFPN) layers for information aggregation and enrichment. The BiFPN introduces top-down feature aggregation (red arrows), bottom-up feature aggregation (green arrows), and feature aggregation from the same scales (blue arrows). Next, a region proposal network (RPN) module and a region of interest (ROI) alignment module are used to generate bounding box proposals based on the BiFPN features. The proposal features are further fed into four convolutional (conv) layers. Finally, two fully-connected layers conduct classification and regression based on the proposals, respectively, and generate the predictions.

*Data Preprocessing*

The original DS1 chest X-ray images were grayscale UNIT16 images in DICOM format. The chest radiographs went through several preprocessing steps before being used to train the DL models. We first calculated the mean and variance of each chest radiograph and clipped the range of intensity values into [mean–3×variance, mean+3×variance] to reduce the outlier points. Each image was then normalized to have zero mean and unit variance. The radiographs from datasets other than DS1 were directly normalized to have an intensity range of zero mean and unit variance.

We concatenated three copies of one chest radiograph to construct three-dimensional inputs for the DL models. For CheXNet, the input chest radiographs were linearly scaled into [0, 1] and resized to 512×512 pixels. For CheXDet, the global mean and variance computed from ImageNet (31) were used for final normalization, and the input images were resized to 768×768 pixels. For data augmentation, we randomly flipped the input images horizontally to enrich the training data.

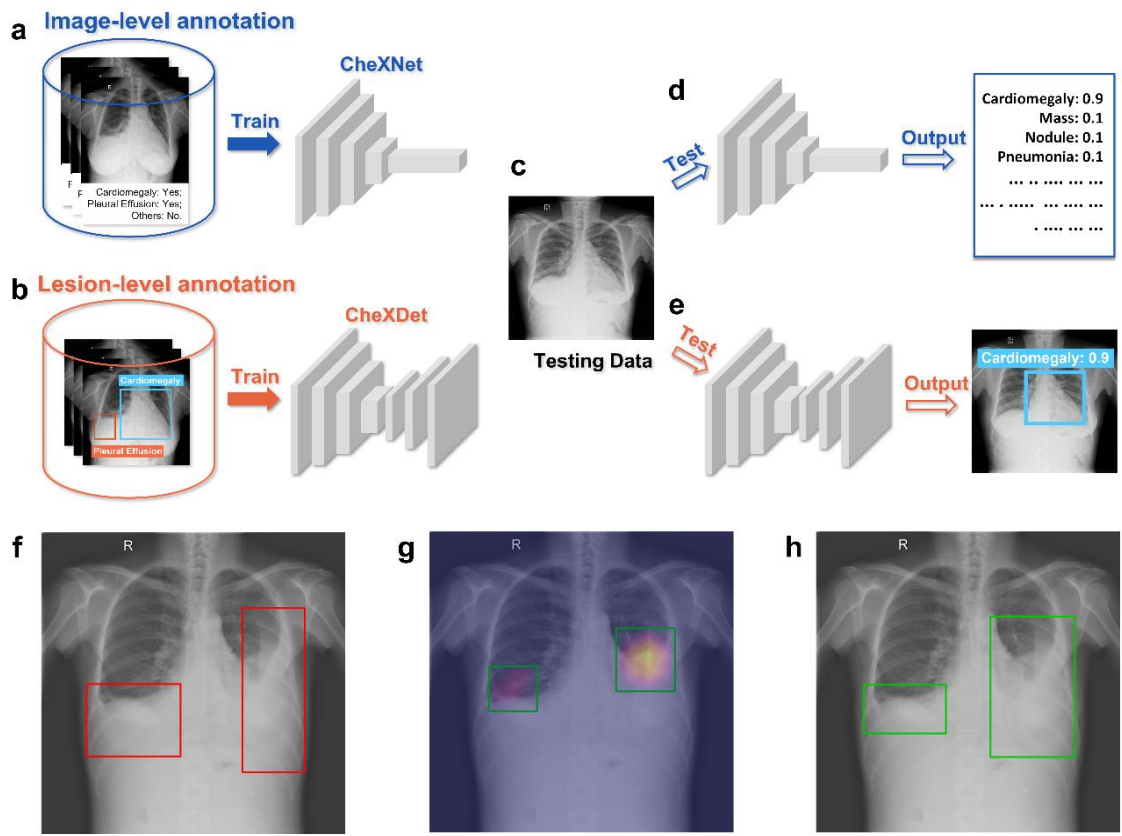

**Figure 3: Training-Testing flows of CheXNet and CheXDet.** (a) CheXNet is trained with radiograph-level annotations that indicate whether specific diseases exist on the whole radiograph. (b) CheXDet is trained with lesion-level annotations that further point out the exact locations of lesions with bounding boxes. (c) A testing chest radiograph. (d) CheXNet predicts the probabilities of each pathology. (e) CheXDet could identify the lesion regions with corresponding disease scores. (f) Another testing image with pleural effusion lesions bounded in the red boxes. (g) The localization results were given by CheXNet using the class activation map method, which is widely adopted by researchers to interpret the results of a deep classification model. Lighter color indicates a higher probability of abnormality found by CheXNet. The top 40% pixels on the heatmaps are bounded by the green boxes as the final localization results. (h) The localization results given by CheXDet indicated with green boxes.

*Model Evaluation and Comparison*

To study shortcut learning and the effect of fine-grained annotations, we evaluated CheXNet and CheXDet performance in two tasks: disease classification and lesion localization.

For the disease classification task, we compared performance on the internal testing set between CheXNet and CheXDet with varying numbers of training data. We also compared CheXNet$_{100+}$ with CheXNet$_{100}$ to validate whether incorporating additional training data improves classification performance. On the testing subset, we compared CheXNet$_{100}$ and CheXDet$_{100}$ with three radiologists (with 4, 13, and 19 years of experience in chest radiology, respectively; neither original chest imager nor involved in the groundtruth labeling). For the lesion detection task, we compared CheXNet[100] with CheXDet[20], CheXDet[40], CheXDet[60], CheXDet[80], and CheXDet[100].

To investigate whether the models could achieve acceptable performance, e.g., comparable performance to radiologists, we further compared CheXNet[100] and CheXDet[100] with three additional radiologists. These radiologists were asked to independently classify the radiographs from the testing subset given only the image data. Readers' performance was reported in sensitivities and specificities for each disease. More details of the reader study process can be found in the appendix (section 1).

If shortcut learning was alleviated, the model would learn more precise features for the diseases, improving their generalizability for external testing. Therefore, disease classification performance of CheXNet[100], CheXNet[100+], CheXDet[20], and CheXDet[100] were evaluated on NIH-Google and PadChest. Note that NIH-Google contains a class "Mass or Nodule", which treats nodule and mass as the same class. We thus took the maximum prediction between the two classes, mass and nodule, to be a single probability for the class "Mass or Nodule". External lesion localization performance of CheXNet[100], CheXDet[20], and CheXDet[100] was evaluated on the NIH ChestX-ray14 dataset. While the external datasets did not cover all diseases studied in DS1, we reported performance on the classes that overlapped with those used in DS1.

*Evaluation Metrics and Statistical Analysis*
To evaluate the disease classification performance, we used the area under the receiver operating characteristic (ROC) curve (AUC). We used the DeLong test (32) to compute the 95% CIs and p-values for the ROC curves. CheXNet generates radiograph-level probabilities for each disease, and the AUCs could hence be directly computed. CheXDet outputs multiple bounding boxes with disease probabilities for each image, and we hence took the maximum probability among every box as the radiograph-level prediction and computed the AUCs.

To evaluate the lesion localization performance, we used the weighted alternative free-response receiver operating characteristics (wAFROC) as the figures of merit (FOMs) by jackknife AFROC (JAFROC) (version 4.2.1; https://github.com/dpc10ster/WindowsJafroc). We performed the 95% CI computation and significance test for JAFROC-FOMs, applying the Dorfman-Berbaum-Metz model with the fixed-case, random-reader method (33). For CheXDet, we filtered out the generated bounding boxes with a threshold where the summation of sensitivity and specificity was the highest, and the remained bounding boxes were used to compute the JAFROC of CheXDet. For CheXNet, we thresholded the heatmaps generated by the gradient-weighted class activation map (Grad-CAM) (34) and obtained the connected components as the detection results. A predicted bounding box would be regarded as a true positive if the intersection-over-union with a groundtruth bounding box was greater than 0.5. The generated bounding boxes were then used to compute the JAFROC of CheXNet. More details for obtaining lesion-level results of CheXNet are in the appendix (section 3.2).

All statistical tests were two-sided. All the measurements and statistical analyses were done using R software (version 3.6.0) (35). We reported p-values after the adjustment with the Benjamini–Hochberg procedure (36) to control the false discovery rate for multiple testing, and we considered a post-adjusted p-value less than .05 to be significant.

Data Availability
The code used in this study can be acquired upon reasonable request from the corresponding author (HC).

*Summary of Datasets*
For DS1, the patients aged $49 \pm$ [standard deviation] 19 years, including 16,959 males, 11,458 females, and 2,144 where sex was unknown. The detailed characteristics are summarized in Table 1.

**Table 1: Clinical Characteristics of Each Dataset**

| | DS1 Training | DS1 Tuning | DS1 Testing | CheXpert Addtional Training | NIH-Google External Testing | PadChest External Testing | NIH ChestX-ray14 External Testing |
|---|---|---|---|---|---|---|---|
| **Data Scale** | | | | | | | |
| **Patients** | 25,019 | 2,751 | 2,791 | 62,170 | 860 | 22,953 | 726 |
| **Images** | 28,673 | 2,906 | 2,922 | 179,919 | 1,962 | 24,536 | 880 |
| **Sex** | | | | | | | |
| **Male** | 13,848 | 1,525 | 1,586 | 34,534 | 490 | 10,716 | 412 |
| **Female** | 9,405 | 1,052 | 1,001 | 27,635 | 370 | 12,235 | 314 |
| **Unknown** | 1,766 | 174 | 204 | 1 | 0 | 2 | 0 |
| **Age** | | | | | | | |
| **Mean (SD)** | 49 (19) | 49 (18) | 49 (18) | 61 (18) | 47 (17) | 59 (18) | 49 (21) |

Note.—The clinical characteristics of the datasets used. DS1 = dataset 1

## Results

*Comparison of Internal Disease Classification Performance on Datset 1*

Figure 4 illustrates the AUCs with 95% CIs of the different models on the internal testing set. AUC, sensitivity, and specificity values of each model can be found in Table E1.

Given the same amount (at least 40%) of training data, there were no statistically significant differences (p-values>.05) between performance of CheXNet and CheXDet. To investigate whether the failure cases of CheXDet were also failures of CheXNet, we compared the false positives (FPs) and false negatives (FNs) of CheXNet$_{100}$ and CheXDet$_{100}$. Among the FPs of CheXDet, for cardiomegaly, effusion, mass, nodule, pneumonia, pneumothorax, tuberculosis, fracture, and aortic calcification, there were 69.4% (154/222), 66.3% (134/202), 40.8% (142/348), 51.5% (234/454), 66.9% (368/550), 55.5% (106/191), 54.4% (160/294), 39.4% (117/297), and 38.9% (82/211) samples that were also the FPs of CheXNet, respectively. Among the FNs of CheXDet, for cardiomegaly, effusion, mass, nodule, pneumonia, pneumothorax, tuberculosis, fracture, and aortic calcification, there were 66.7% (8/12), 57.1% (32/56), 60.6% (20/33), 50.0% (46/92), 54.9% (39/71), 55.0% (11/20), 72.3% (68/94), 25.0% (3/12), and 77.8% (7/9) samples that were also the FNs of CheXNet, respectively.

With only 20% of the training data, CheXNet$_{20}$ achieved higher performance than CheXDet$_{20}$ in classifying pneumonia (AUC: 0.85 vs 0.82; p-value<.05), tuberculosis (AUC: 0.91 vs 0.90, p-value<.05), and fracture (AUC: 0.91 vs 0.86, p-value<.05), but lower performance than CheXDet$_{20}$ in classifying pneumothorax (AUC: 0.92 vs 0.95, p-value<.05) and aortic calcification (AUC: 0.85 vs 0.94, p-value<.001).

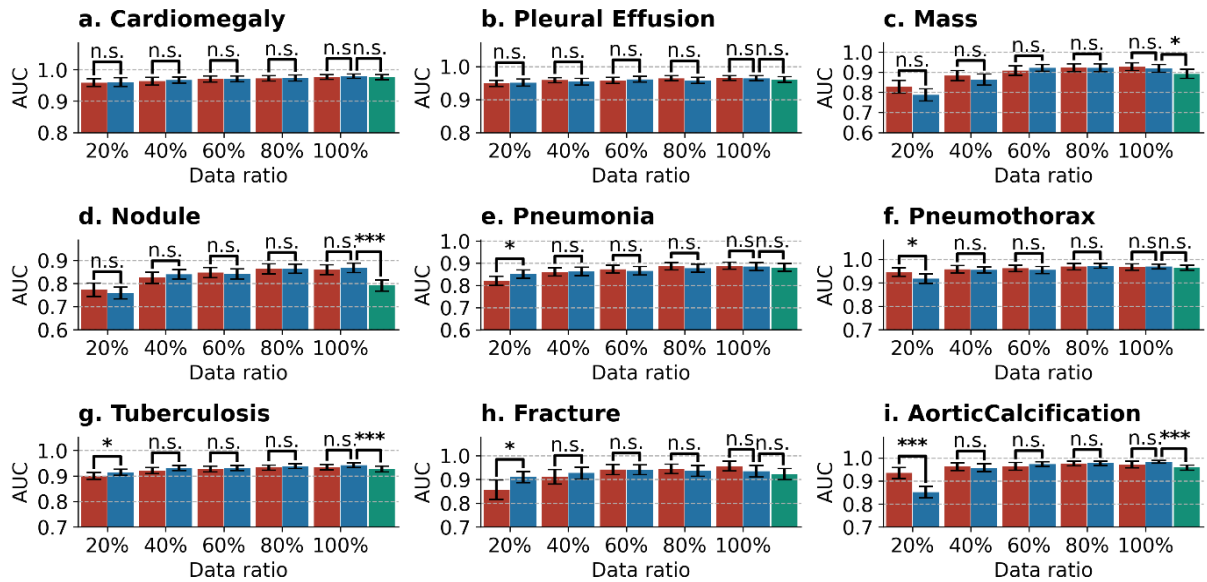

**Figure 4: Disease classification performance of models on the internal testing set under different ratios of training data.** Blue bars represent the area under the receiver operating characteristic curves (AUCs) with 95% CIs for CheXNet; red bars represent AUCs with 95% CIs for CheXDet; and green bars represent AUCs with 95% CIs for CheXNet trained with additional data from CheXpert dataset. Under many scenarios, CheXDet and CheXNet achieve similar performance without evidence of a difference on the internal disease classification task. Whiskers represent the CIs. n.s.: not significant. *p < .05. ***p<.001.

CheXNet$_{100+}$ trained with 100% of DS1 plus additional CheXpert data showed lower performance in classifying four out of nine diseases than CheXNet$_{100}$, including mass (p-value<.05), nodule (p-value<.001), tuberculosis (p-value<.001), and aortic calcification (p-value<.001). In classifying the other five diseases, CheXNet$_{100+}$ showed no evidence of a difference compared with CheXNet$_{100}$.

*Comparison of Deep Learning Models with Radiologists on Dataset 1*

Figure 5 illustrates the performance of the three radiologists and the ROC curves of CheXNet$_{100}$ and CheXDet$_{100}$ on the testing subset. The radiologists showed high specificities for classifying all the diseases with trade-offs on sensitivities. All the points representing individual experts lie on or near to the right of the ROC curves of the models, indicating thresholds where the models performed on par or better than radiologists.

*Comparison of External Classification Performance on NIH-Google*

Table 2 reports the AUCs with CIs for CheXNet$_{100}$, CheXNet$_{100+}$, CheXDet$_{20}$, and CheXDet$_{100}$, as well as the p-values for comparisons with CheXNet$_{100}$ on NIH-Google. CheXNet$_{100+}$ showed considerably lower performance than CheXNet$_{100}$ on classifying nodule or mass (p-value<.001) on this external set; whereas there was no evidence of a difference for pneumothorax and fracture classification between these two models. CheXDet$_{20}$ and CheXDet$_{100}$ achieved higher performance than CheXNet$_{100}$ on classifying nodule or mass (p-values<.001) and fracture (p-values<.001), and CheXDet$_{100}$ also showed higher AUC on classifying pneumothorax (p-value<.05) than CheXNet$_{100}$.

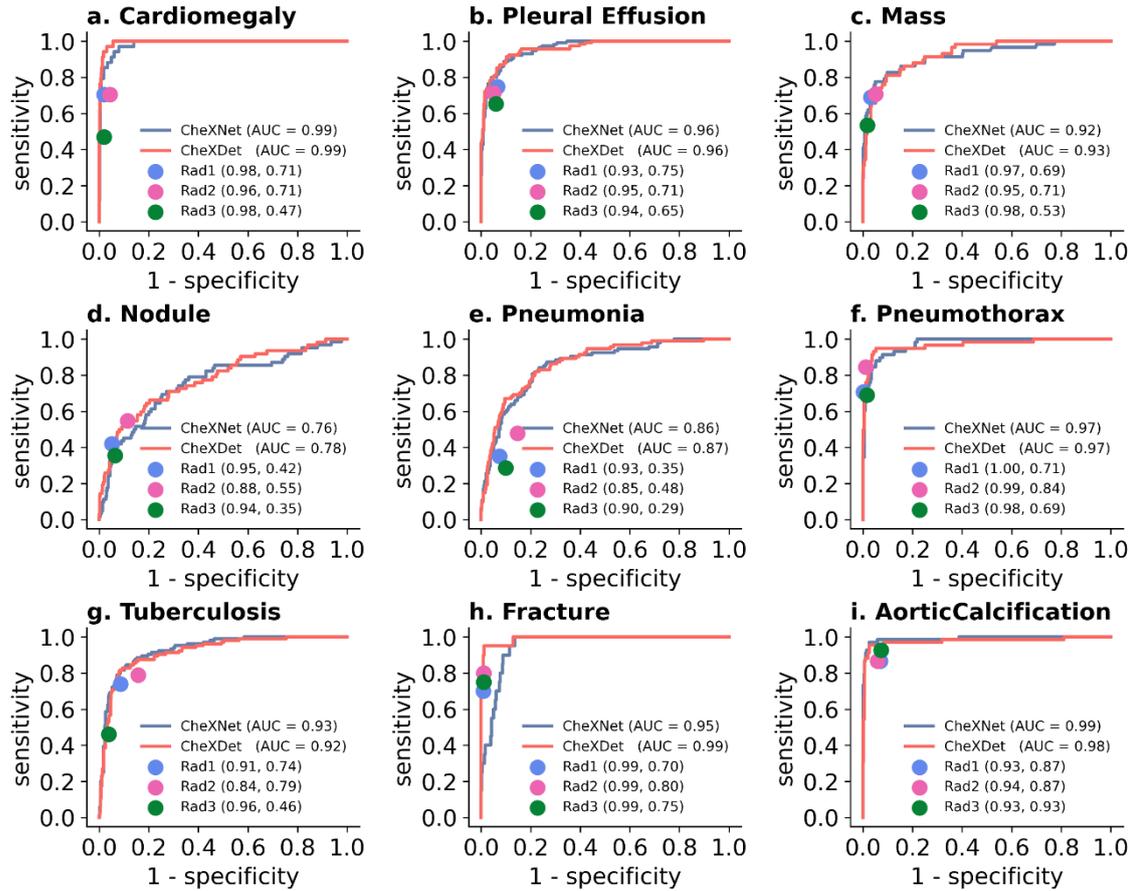

**Figure 5: Comparison of radiograph-level pathology classification performance among models and radiologists on the testing subset.** Blue curves represent receiver operating characteristic curves (ROCs) of CheXNet. Red curves represent ROCs of CheXDet. Radiologists' (Rad) performance levels are represented as single points. Radiologist performance is reported in parentheses as (specificity, sensitivity). Almost all the points representing individual radiologists lie on or under the ROC of one of the models, which means there exist thresholds where at least one model performs on par or better than practicing radiologists. AUC = area under the ROC curve.

*Comparison of External Classification Performance on PadChest*

Table 2 also reports the AUCs with CIs for CheXNet$_{100}$, CheXNet$_{100+}$, CheXDet$_{20}$, and CheXDet$_{100}$, as well as p-values for comparisons with CheXNet$_{100}$ on PadChest.

CheXNet$_{100+}$ showed higher performance on cardiomegaly (p-value<.001), mass (p-value<.001), and fracture classification (p-value<.05) but lower performance on nodule (p-value<.001), pneumonia (p-value<.01), and aortic calcification (p-value<.001) than those of CheXNet$_{100}$. These was no evidence of differences between these two models in classifying pleural effusion, pneumothorax, and tuberculosis.

Meanwhile, CheXDet$_{20}$ achieved higher performance than CheXNet$_{100}$ on mass (p-value<.001), nodule (p-value<.001), and fracture classification (p-value<.001) and lower performance on classification of cardiomegaly (p-value<.001), pleural effusion (p-value<.01), and aortic calcification (p-value<.05). There was no evidence of a difference in model performance for classifying pneumonia, pneumothorax, and tuberculosis.

**Table 2: Comparison of Chest Radiograph Classification Performance Between Models on the External Datasets**

| Dataset | Diseases | CheXNet₁₀₀ | CheXNet₁₀₀₊ | | CheXDet₂₀ | | CheXDet₁₀₀ | |
|---|---|---|---|---|---|---|---|---|
| | | AUC (95% CI) | AUC (95% CI) | p-value | AUC (95% CI) | p-value | AUC (95% CI) | p-value |
| **NIH-Google** | Nodule or Mass | 0.68 (0.66, 0.70) | 0.64 (0.61, 0.66) | .002 | 0.74 (0.72, 0.76) | < .001 | *0.80 (0.78, 0.81) | < .001 |
| | Pneumothorax | 0.84 (0.81, 0.87) | 0.84 (0.82, 0.87) | .92 | 0.82 (0.80, 0.85) | 0.22 | *0.87 (0.85, 0.89) | .03 |
| | Fracture | 0.51 (0.47, 0.55) | 0.51 (0.46, 0.55) | .92 | 0.66 (0.62, 0.70) | < .001 | *0.67 (0.63, 0.71) | < .001 |
| **PadChest** | Cardiomegaly | 0.91 (0.91, 0.92) | *0.92 (0.92, 0.93) | < .001 | 0.88 (0.88, 0.89) | < .001 | 0.91 (0.91, 0.92) | .61 |
| | Pleural Effusion | *0.95 (0.94, 0.96) | *0.95 (0.94, 0.96) | .91 | 0.94 (0.93, 0.95) | .007 | 0.94 (0.93, 0.95) | .06 |
| | Mass | 0.55 (0.53, 0.57) | 0.59 (0.57, 0.61) | < .001 | 0.67 (0.65, 0.69) | < .001 | *0.63 (0.61, 0.65) | < .001 |
| | Nodule | 0.66 (0.63, 0.69) | 0.55 (0.53, 0.58) | < .001 | 0.73 (0.70, 0.75) | < .001 | *0.78 (0.76, 0.80) | < .001 |
| | Pneumonia | 0.79 (0.77, 0.81) | 0.77 (0.76, 0.79) | .002 | 0.80 (0.79, 0.82) | .11 | *0.83 (0.81, 0.84) | < .001 |
| | Pneumothorax | 0.83 (0.77, 0.88) | 0.81 (0.75, 0.87) | .62 | 0.78 (0.71, 0.85) | .25 | *0.85 (0.79, 0.92) | .34 |
| | Tuberculosis | 0.89 (0.86, 0.93) | 0.88 (0.85, 0.91) | .60 | 0.90 (0.87, 0.93) | .45 | *0.92 (0.89, 0.95) | .06 |
| | Fracture | 0.55 (0.53, 0.57) | 0.58 (0.56, 0.60) | .01 | 0.74 (0.71, 0.76) | < .001 | *0.78 (0.76, 0.80) | < .001 |
| | Aortic Calcification | 0.86 (0.85, 0.87) | 0.81 (0.79, 0.82) | < .001 | 0.85 (0.84, 0.86) | .04 | *0.87 (0.86, 0.88) | .14 |

Note.—CheXNet, developed with 100% of the dataset 1 (DS1) training data (CheXNet100), was compared against CheXNet trained with 100% of DS1 and additional data from CheXPert (CheXNet100+) and CheXDet developed with 20% (CheXDet20) and 100% of the DS1 training data (CheXDet100) for the radiograph classification performances on the external NIH-Google and PadChest datasets. Best performance is highlighted with *, and p-values were computed between CheXNet100 and every other model. AUC: area under the receiver operating characteristic curve.

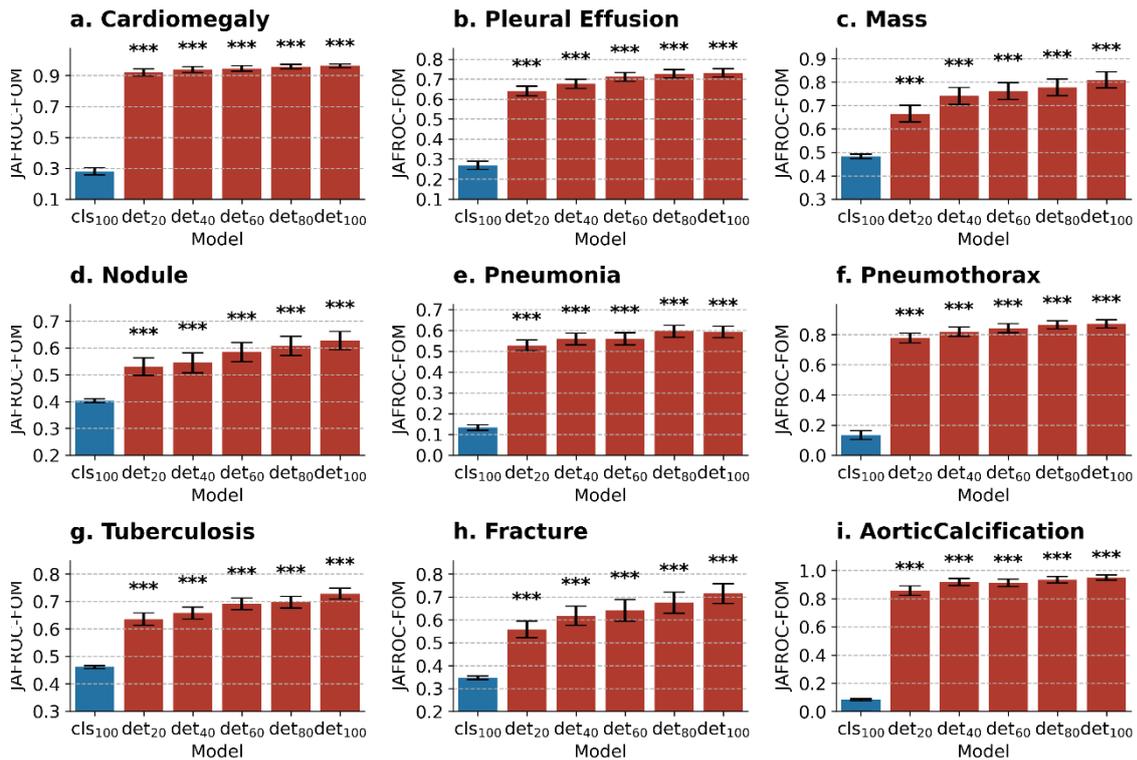

**Figure 6: Comparison of lesion detection performance among models on the internal testing set.** CheXNet developed with 100% data (cls$_{100}$) is compared against CheXDet developed with different ratios of data (det$_{20}$, det$_{40}$, det$_{60}$, det$_{80}$, and det$_{100}$; subscripts denote ratios of training data). Blue bars represent jackknife alternative free-response receiver-operating (JAFROC) figure of merits (FOMs) with 95% CIs of CheXNet, and red bars represent JAFROC-FOMs with 95% CIs of CheXDet. Whiskers represent the CIs. CheXDet performs higher than CheXNet on the internal lesion detection task, even when trained with 20% of the data. ***p < .001.

With the same amount of training data, CheXDet$_{100}$ achieved higher AUCs than CheXNet$_{100}$ in classifying four out of nine diseases, including mass (p-value<.001), nodule (p-value<.001), pneumonia (p-value<.001), and fracture (p-value<.001). Meanwhile, CheXDet$_{100}$ and CheXNet$_{100}$ demonstrated no evidence of a difference in cardiomegaly, pleural effusion, pneumothorax, tuberculosis, and aortic calcification classification.

*Comparison of Internal Lesion Detection Performance on Dataset 1*
Figure 6 illustrates the JAFROC-FOMs with 95% CIs for CheXNet$_{100}$, CheXDet$_{20}$, CheXDet$_{40}$, CheXDet$_{60}$, CheXDet$_{80}$, and CheXDet$_{100}$ as well as p-values compared against CheXNet$_{100}$ on the internal testing set. Here, we compared CheXDet developed with 20%, 40%, 60%, 80%, and 100% training data against CheXNet trained with 100% data. The JAFROC-FOMs of CheXDet on each disease increased progressively by about 10% when the amount of training data increased from 20% to 100%. In all scenarios and for all diseases, CheXDet achieved higher performance (p-values<.001) than CheXNet, even when developed with only 20% of training data. Specific statistics of JAFROC-FOMs with CIs of different models can be found in Table E2.

*Comparison of External Lesion Detection Performance on NIH ChestX-ray14*
Table 3 shows the JAFROC-FOMs with 95% CIs for CheXNet$_{100}$, CheXDet$_{20}$, and CheXDet$_{100}$, as well as p-values compared against CheXNet$_{100}$ on NIH ChestX-ray14. CheXDet$_{20}$ trained with only

**Table 3: Comparison of Lesion Localization Performance Between Models on the External NIH-ChestX-ray14 Testing set**

| | CheXNet$_{100}$ | CheXDet$_{20}$ | | CheXDet$_{100}$ | |
|---|---|---|---|---|---|
| **Disease** | **JAFROC-FOM (95% CI)** | **JAFROC-FOM (95% CI)** | **p-value** | **JAFROC-FOM (95% CI)** | **p-value** |
| **Cardiomegaly** | 0.08 (0.06, 0.11) | 0.65 (0.60, 0.70) | < .001 | *0.79 (0.75, 0.83) | < .001 |
| **Pleural Effusion** | 0.17 (0.12, 0.22) | 0.25 (0.21, 0.29) | .01 | *0.26 (0.21, 0.30) | .002 |
| **Nodule** | *0.31 (0.29, 0.33) | 0.31 (0.25, 0.37) | .99 | 0.30 (0.24, 0.37) | .91 |
| **Mass** | 0.45 (0.44, 0.46) | 0.40 (0.35, 0.46) | .13 | *0.56 (0.48, 0.63) | .009 |
| **Pneumonia** | 0.18 (0.14, 0.22) | 0.33 (0.28, 0.39) | < .001 | *0.34 (0.27, 0.40) | < .001 |
| **Pneumothorax** | 0.04 (0.01, 0.08) | 0.41 (0.34, 0.47) | < .001 | *0.55 (0.48, 0.61) | < .001 |

Note.—CheXNet developed with 100% of DS1 training data (CheXNet100) was compared against CheXDet developed with 20% of DS1 training data (CheXDet20) and 100% of DS1 training data (CheXDet100) for the lesion localization performances on the external NIH ChestX-ray14 dataset. Best performance is highlighted with *, and p-values were computed between CheXNet100 and every other model. JAFROC-FOM: free-response receiver-operating characteristic figure of merit.

20% data also showed apparently higher JAFROC-FOMs than CheXNet$_{100}$ on four out of six diseases, including cardiomegaly (p-value<.001), pleural effusion (p-value<.05), pneumonia (p-value<.001), and pneumothorax (p-value<.001). There was no evidence of differences between CheXDet$_{20}$ and CheXNet$_{100}$ on localizing nodule and mass.

With increased training data, CheXDet$_{100}$ achieved higher performance on five out of six types of lesions, including cardiomegaly (p-value < .001), pleural effusion (p-value < .01), mass (p-value < .01), pneumonia (p-value < .001), and pneumothorax (p-value < .001) than CheXNet$_{100}$, with no evidence of a difference between CheXDet$_{100}$ and CheXNet$_{100}$ on nodule detection.

**Discussion**

In this study, we developed CheXNet and CheXDet and focused on evaluating the models from two aspects following the recommended shortcut learning evaluation practice (16, 17): whether the models attend to the lesion regions and whether the models generalize well for external testing. Existing works have reported shortcut learning in medical image diagnosis AIs (13, 20), yet few of them have tried to quantify or tackle this challenge. We provided a possible solution to make DL models right for the right reasons, which could significantly improve external performance.

Our study showed that for internal testing, incorporating additional training data from CheXpert led to a considerable performance drop on four out of nine diseases. One possible reason is that CheXpert dominated the training set and made CheXNet$_{100+}$ fit on a different distribution from the original distribution of DS1, as the CheXpert dataset is labeled by natural language processing and has a much older participant pool. These observations suggest that incorporating more training data does not always benefit the classification accuracy for DL models and alternative solutions should be sought. On the other hand, CheXDet mainly showed no evidence of differences in internal disease classification

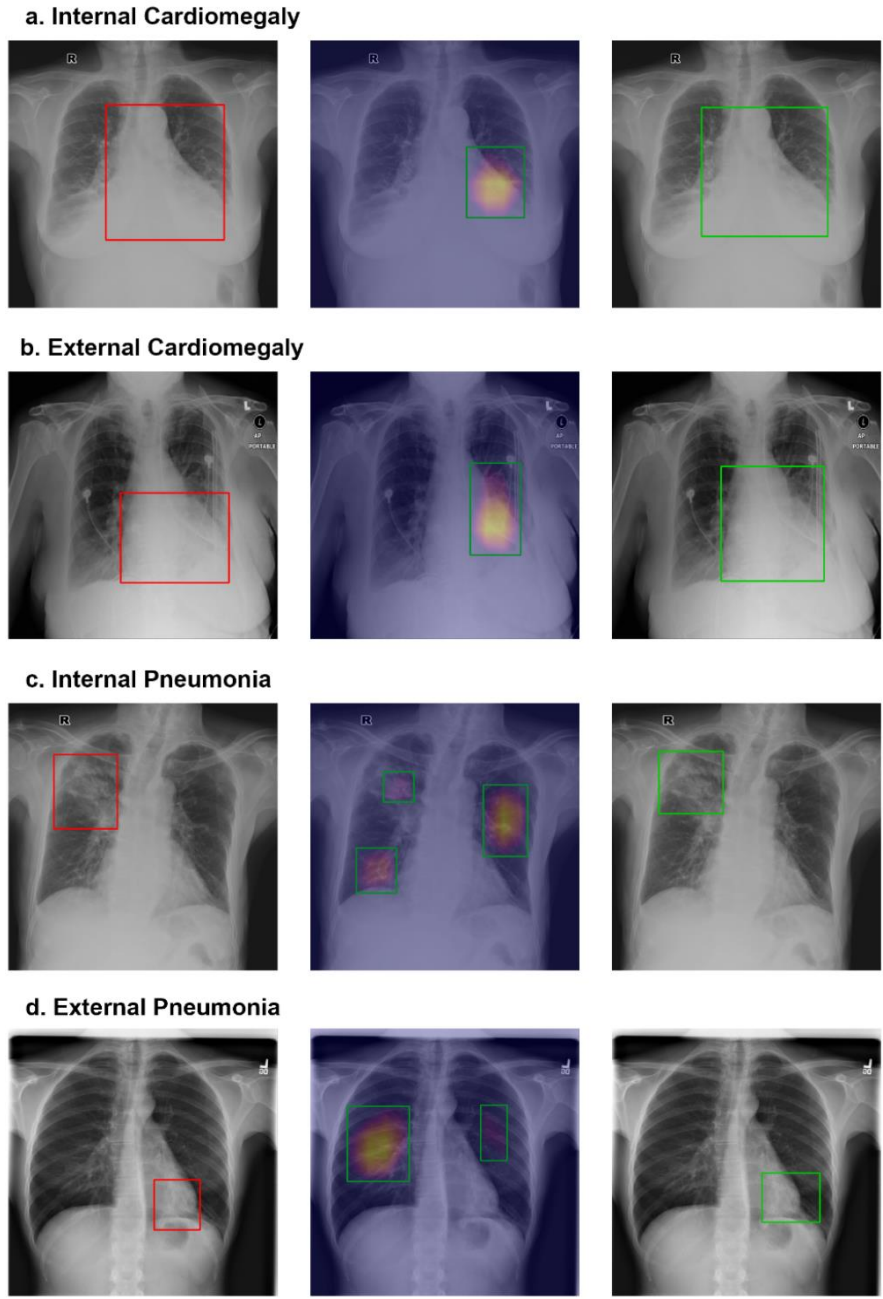

**Figure 7: Sample localization results of CheXNet and CheXDet.** Qualitative samples of lesion localization results for (a-b) cardiomegaly and (c-d) pneumonia on the (a, c) internaldataset, Dataset 1, and (b,d) external set, NIH ChestX-ray14. Groundtruth with bounding boxes (red, left column), gradient-weighted class activation map (Grad-CAM) generated by ChexNet (middle column), and localization of output of CheXDet (right column) are demonstrated. For Grad-CAMs, the color overlay indicates a higher probability of abnormality found by CheXNet, and the top 40% pixels on the heatmaps are bounded by the green boxes as the final localization results. CheXDet outputs the correct bounding box for cardiomegaly, outlining the entirety of the heart (same as groundtruth labeling), while CheXNet only focuses on the left side of the heart. This was evident on both the internal and external datasets. CheXDet localizes the correct locations for pneumonia changes; whereas, CheXNet included non-targeted areas in (c), likely due to fibrotic changes, and missed the targeted area entirely in (d), instead identifying false positive areas which appear normal radiographically.

compared with CheXNet when given the same amount (at least 40%) of training data from DS1.

Generalizability on external datasets is crucial for determining whether a DL system can be applied to real-world clinical usage. Existing works have proposed solutions that focused on increasing the diversity of the training data (e.g., with data augmentation techniques or training with multicenter data) (37). Here, we demonstrated that fine-grained annotations significantly improved the DL model's generalizability on chest radiographs from new centers, (i.e., NIH-Google and PadChest), without training the models with multicenter data. Specifically, CheXDet could achieve higher external performance than $CheXNet_{100+}$ without loss of accuracies on the internal data. When both were developed by all training data from DS1, $CheXDet_{100}$ also outperformed $CheXNet_{100}$ for all three diseases on NIH-Google and four out of nine diseases on PadChest without degraded performance on other diseases. Moreover, for small lesions without fixed positions such as nodules, masses, and fractures, even $CheXDet_{20}$ performed higher than $CheXNet_{100}$ despite being developed with only 20% of the training data. Importantly, these findings suggest that DL models developed with fine-grained lesion annotations are more generalizable to external data.

As Grad-CAM has been widely adopted in many previous works to show that CheXNet could identify correct disease signs, we quantified the disease localization capability of CheXNet and CheXDet. Our data revealed that CheXNet relies highly on patterns other than the true pathological signs to make decisions, as it showed low performance in finding the lesions but achieved radiologist-level internal classification results. Apart from providing quantitative comparison, we present some sample detection results of $CheXNet_{100}$ and $CheXDet_{100}$ in Figure 7. For internal data, it can be observed that CheXNet's Grad-CAM might not precisely cover the intended lesions (Figure 7a) and sometimes even attend to false-positive regions (Figure 7c). Moreover, CheXNet might then use incorrect patterns to make decisions for the external data (Figure 7b and Figure 7d). The degraded lesion detection performance and external classification performance of CheXNet, together with the visualization results, demonstrated that a DL model trained with radiograph-level annotations is prone to shortcut learning, i.e., using unintended patterns for decision-making. Worse yet, such a model achieved comparable performance to radiologists on the internal testing subset, as shown in Figure 5. Based on these results, the claim that deep learning demonstrates comparable performance to doctors may need further investigation. On the contrary, training with fine-grained annotations enabled CheXDet to focus on the correct pathological patterns and become more robust to external data and less prone to shortcut learning. Our findings highlight the importance of using fine-grained annotations for developing trustworthy DL-based medical image diagnoses.

We acknowledge the limitations of the current work. First, we chose NIH-Google, PadChest, and NIH ChestX-ray14 as the external testing sets, which were the few publicly available datasets hand-labeled by radiologists. As some external testing datasets did not obtain the same disease categories as our internal dataset, we could only test the diseases that overlapped with our annotations. Second, based on recent studies (38, 39), developing a localization model does not completely address automatic radiograph screening. CheXDet also had failure cases, as shown in the examples from Figure E3. Moreover, our results showed that the external performance of CheXDet was not as good as its internal performance. As this performance drop could imply shortcut learning (16), we believe that CheXDet also experienced shortcut learning but to a lesser degree than CheXNet. Third, fine-grained annotations bring more burden on the labelers. The trade-off between the labor for fine-grained annotations and the improved generalizability thus remains to be explored and elaborated.

To summarize, we showed that a DL model trained with radiograph-level annotations was prone to shortcut learning that used unintended patterns for decision-making for disease detection on chest radiographs. We also showed that fine-grained annotations on chest radiographs improve DL model-based diagnosis, especially when applied to external data, by alleviating shortcut learning and correcting the decision-making regions for the models. We highlighted that successful application of AI models to clinical use lies in the annotation granularity in addition to data size and model architecture, which requires further investigation.

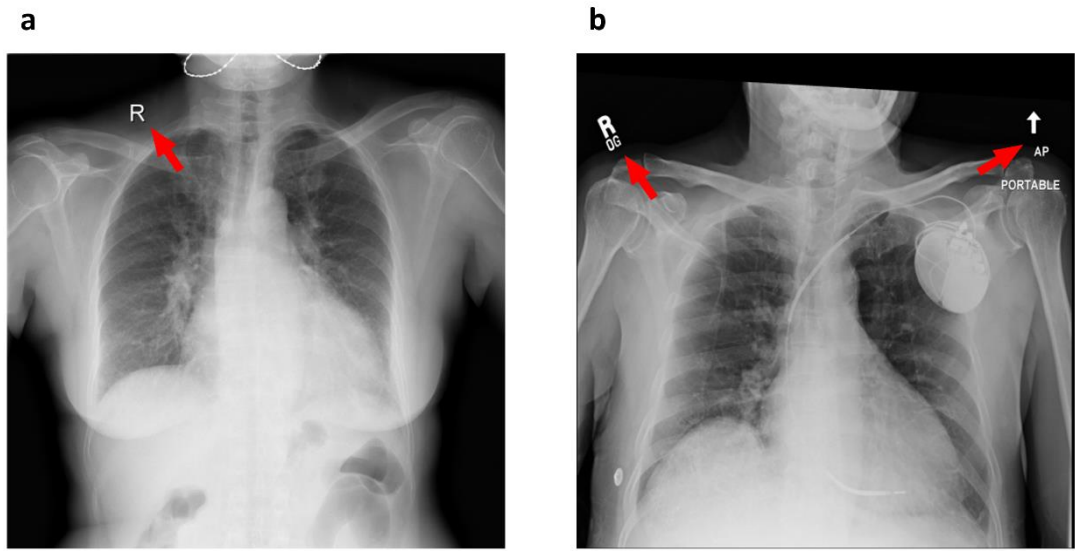

**Figure E1: Sample radiographs with hospital tokens (highlighted by red arrows).** Some previous studies found that the hospital tokens could be used as cues for making decisions when detecting abnormalities from chest radiographs, which are typical examples of shortcut learning.

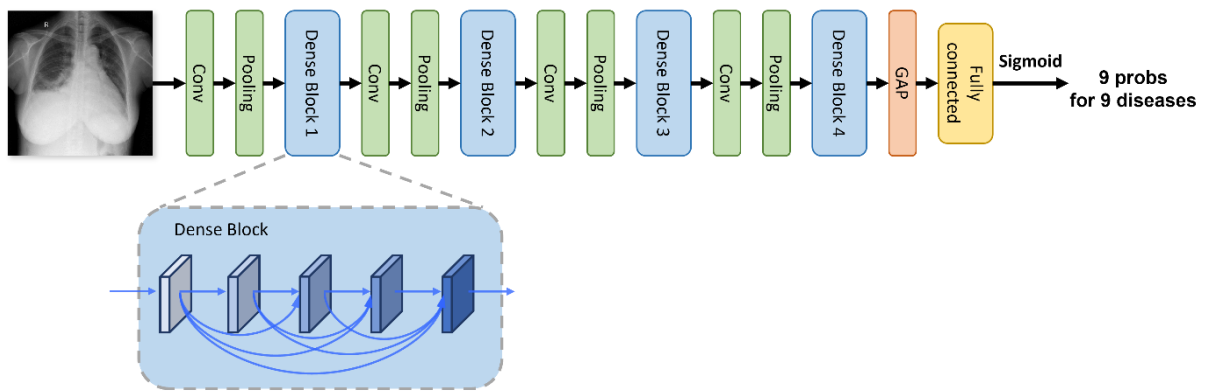

**Figure E2: The brief architecture of CheXNet.** CheXNet is a 121-layer densely-connected convolutional neural network with the outputs of probabilities for nine pathologies.

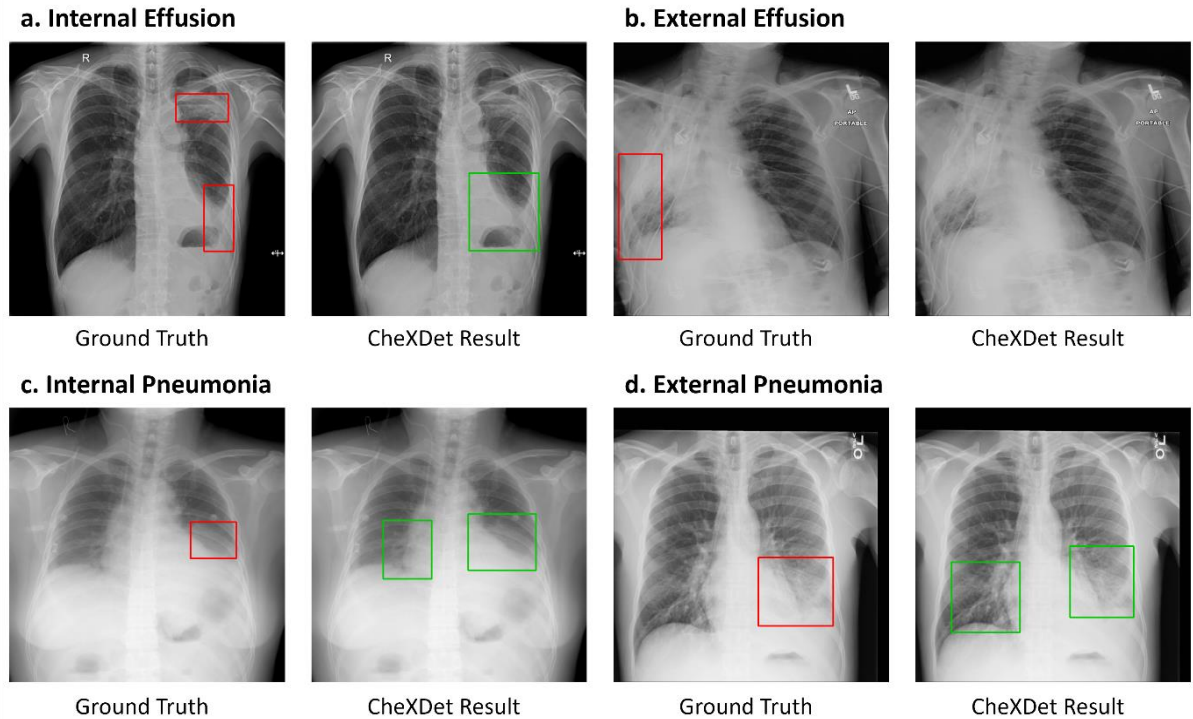

**Figure E3: Failure cases of CheXDet.** CheXDet produced false negatives when the radiographs appeared to be (a) overexposed or (b) of low signal-to-noise ratio. For pneumonia, CheXDet produced false positives at the overlay area of the right lung and heart, as shown in both (c) and (d). Red boxes represent the ground truth bounding boxes, and green boxes represent the predicted bounding boxes given by the model.

**Supplementary Materials**

**1. Reader study**

The three radiologists that compared with the models had 4, 13, and 19 years of experience in chest radiology, respectively. They were not the original chest imager, nor have they participated in the groundtruth labeling process. The readers were provided with the same software used in the groundtruth labeling. All images were of the same size as their original digital imaging and communications in medicine (DICOM) format, but other information in the DICOM was not provided. They made binary decisions on the available frontal view radiographs, blinded to the radiology text reports and other readers' annotations. All readers were provided with the same guidelines for the annotation software and criteria.

**2. Development of CheXNet**

**2.1. Architecture and training of CheXNet**

We developed CheXNet following Rajpurkar et al. (1), except that we required the model to output probabilities for nine pathologies. The model was a 121-layer densely-connected network (DenseNet) (2) pre-trained on the ImageNet dataset. The DenseNet contained four dense blocks, where the features from every shallow layer were concatenated and fed into the deeper layers to enable better gradient back-propagation. A convolutional layer and a pooling layer were appended after each dense block to conduct dimension reduction. We replaced the original final output layer (1000-way softmax) of this network with nine-way sigmoid. During training, we randomly flipped the input images horizontally to enrich the training data. We set the training objective as minimizing the average per-class cross-entropy between the model's prediction and the groundtruth (0 for absence and 1 for presence of a specific disease). The training objective was formed by multiple binary classification losses:

$$L_{\text{BCE}} = \sum_{i=1}^{N} (-y_i \log(p_i) - (1 - y_i) \log(p_i))$$

where $L_{\text{BCE}}$ stands for binary cross-entropy loss, $y$ is the groundtruth label, $p$ is the predicted probability, $N$ is the number of classes (diseases), and $i$ is the index for a specific class. The constructed network is depicted in Figure E2.

When training CheXNet with additional data from CheXpert, we only used diseases labeled in CheXpert that were overlapped with those in DS1. We used Adam (3) as the optimization solver, with $\beta_1$ and $\beta_1$ was set to 0.9 and 0.999, respectively, as recommended in the original Adam optimizer paper. We set the initial learning rate to 0.001 and control the change of it by a cosine annealing strategy as follows:

$$\alpha = 0.5 \times \alpha_0 \times (1 + \cos(\frac{\pi \times t}{T})) \ ,$$

where $\alpha$ is the learning rate at epoch t, $\alpha_0$ is the initial learning rate, and T is the maximum epoch number (set as 25). We input 48 radiographs to the network for each model updating step until the whole training set had been fed to the model 25 epochs. We evaluated the model on the tuning set after each epoch and chose the one with the best AUCs for subsequent experiments on the testing set. The model was trained on 4 TITAN XP GPUs with 12 GB of memory each. The implementations were done using PyTorch (https://pytorch.org) deep learning framework.

## 2.2 Lesion localization results by CheXNet

We used gradient-weighted class activation mapping (Grad-CAM) (4) to generate disease heatmaps from CheXNet. Higher intensities on the heatmaps indicated higher probabilities for suspicious lesions. The heatmaps were linearly rescaled to be in the range [0, 1]. Next, we applied thresholding on the heatmaps to obtain the regions with top 40% intensities (i.e., normalized intensity > 0.6). Then, bounding boxes were generated for each connected component. The product of the whole radiograph-level probability and the maximum rescaled intensity inside the bounding box was used as the lesion-level probability.

## 3. Development of CheXDet

### 3.1 Architecture and training of CheXDet

CheXDet was required to predict proposals of possible lesion regions as well as the corresponding pathology probabilities. The feature extractor backbone, EfficientNet[7], was a convolutional neural network initially designed by applying the neural network architecture search technique on the ImageNet dataset. The backbone down-sampled the input gradually to 1/64 with a scale of two, which could be divided into five stages accordingly. The features generated at stages 2, 3, 4, 5, and 6 were fed into three bi-directional feature pyramid network (BiFPN) (5) layers. The BiFPN module introduced cross-scale feature aggregation to enrich the information sharing among features of different scales. As shown in Figure 2, the BiFPN introduces top-down feature aggregation (red arrows), bottom-up feature aggregation (green arrows), and feature aggregation from the same scales (blue arrows). The following equation is used for the feature aggregation:

$$O = \sum_i \frac{w_i}{\epsilon + \sum_j w_j} \cdot I_i,$$

Where $O$ is the output for feature aggregation, $I_i$ is the input feature, $w_i$ is a learnable weighting parameter, and $\epsilon$ is a small number used to avoid zero-division. Then, the outputs of BiFPN were fed into a region proposal network (RPN) (6) which generated proposal boxes and predicted them to be foregrounds (i.e., lesions) or backgrounds (i.e., healthy tissues). The loss functions in RPN contain a binary cross-entropy loss for foreground-background classification and a smooth L1 loss for box regression. In specific, the classification loss is as follows:

$$L_{\text{cls}} = -y\log(p) - (1 - y)\log(p),$$

where $L_{\text{cls}}$ stands for the region-wise (bounding box-wise) classification loss, $y$ denotes binary label for foreground or background, and $p$ is the region probability. The smooth L1 loss is as follows:

$$L_{\text{reg}} = \begin{cases} 0.5 \times (t - x)^2, & if \ |t - x| < 1 \\ |t - x| - 0.5, & otherwise \end{cases},$$

where $L_{\text{reg}}$ stands for the smooth L1 loss, $t$ is the transformed coordinates of groundtruth bounding boxes, and $x$ is the transformed coordinates of bounding box predictions.

Then, the generated proposals with the highest foreground scores would be further fed into an ROI (region of interest) alignment module (7) for proposal alignment. The proposal features were further fed into four convolutional layers, and finally two fully-connected layers were used to predict the specific abnormality categories and bounding-box locations. The loss function contains a multi-class cross-entropy loss for disease category classification and smooth L1 loss for final box refinement. The smooth L1 loss is

the same as aforementioned, and the multi-class cross-entropy loss for each proposal is as follows:

$$L'_{cls} = \frac{1}{N} \sum_i^N -c_i \log(q_i),$$

where $L'_{cls}$ is the multi-class cross-entropy loss, $c_i$ is the binary label for the $i^{th}$ category (disease), and $q_i$ is the prediction indicating the probability of the proposal for the $i^{th}$ category (disease). The lesion for the $i^{th}$ disease is present inside the bounding box if $c_i = 1$ and absent otherwise. The constructed network is depicted in Figure 2.

We initialized the CheXDet feature extraction backbone with the pre-trained weights of EfficientNet from ImageNet (8) and reset the output size of the network to be nine classes. We also randomly flipped the input images horizontally to enrich the training data. We used Stochastic Gradient Descent (SGD) with a momentum of 0.9 and a weight decay of 0.0005. We fed each time three images to the network and trained the model for 120,000 iterations. The learning rate was initially set to 0.005 and decayed twice with a ratio of 0.1 at steps 70,000 and 100,000, respectively. We evaluated the model every 5000 steps on the tuning set and chose the checkpoint with the best lesion-level accuracy (mean average precision over nine diseases) for later-on experiments on the testing set. The models were trained on one TITAN XP GPU with 12 GB memory. The implementations were done using the Detectron2 (https://github.com/facebookresearch/detectron2) package based on the PyTorch (https://pytorch.org) framework.

**Table E1. Detailed AUCs, sensitivities and specificities of CheXNet and CheXDet on DS1.**

| Disease | Ratio of training data | CheXNet | | | CheXDet | | |
|---|---|---|---|---|---|---|---|
| | | AUC (95% CI) | Sensitivity (95% CI) | Specificity (95% CI) | AUC (95% CI) | Sensitivity (95% CI) | Specificity (95% CI) |
| Cardiomegaly | 20% | 0.96 (0.95, 0.98) | 0.92 (0.87, 0.96) | 0.91 (0.84, 0.94) | 0.96 (0.95, 0.97) | 0.92 (0.87, 0.96) | 0.89 (0.80, 0.92) |
| | 40% | 0.97 (0.96, 0.98) | 0.89 (0.85, 0.94) | 0.93 (0.88, 0.95) | 0.96 (0.95, 0.98) | 0.93 (0.88, 0.96) | 0.88 (0.83, 0.91) |
| | 60% | 0.97 (0.96, 0.98) | 0.95 (0.91, 0.98) | 0.90 (0.72, 0.92) | 0.97 (0.96, 0.98) | 0.93 (0.88, 0.96) | 0.91 (0.85, 0.94) |
| | 80% | 0.97 (0.96, 0.98) | 0.96 (0.92, 0.98) | 0.88 (0.74, 0.90) | 0.97 (0.96, 0.98) | 0.92 (0.87, 0.96) | 0.92 (0.88, 0.95) |
| | 100% | 0.98 (0.97, 0.99) | 0.95 (0.91, 0.98) | 0.91 (0.85, 0.93) | 0.98 (0.97, 0.98) | 0.94 (0.91, 0.98) | 0.92 (0.86, 0.94) |
| | 100% + CheXpert | 0.98 (0.97, 0.98) | 0.96 (0.92, 0.98) | 0.90 (0.80, 0.93) | NA | NA | NA |
| Pleural Effusion | 20% | 0.95 (0.94, 0.96) | 0.91 (0.88, 0.94) | 0.86 (0.82, 0.89) | 0.95 (0.94, 0.96) | 0.89 (0.86, 0.92) | 0.88 (0.84, 0.90) |
| | 40% | 0.95 (0.94, 0.96) | 0.90 (0.86, 0.92) | 0.89 (0.85, 0.91) | 0.96 (0.95, 0.97) | 0.92 (0.89, 0.95) | 0.85 (0.82, 0.88) |
| | 60% | 0.96 (0.95, 0.97) | 0.90 (0.86, 0.93) | 0.90 (0.86, 0.93) | 0.96 (0.95, 0.97) | 0.88 (0.84, 0.91) | 0.90 (0.86, 0.93) |
| | 80% | 0.96 (0.95, 0.97) | 0.91 (0.88, 0.94) | 0.89 (0.85, 0.91) | 0.96 (0.96, 0.97) | 0.89 (0.85, 0.92) | 0.91 (0.87, 0.94) |
| | 100% | 0.96 (0.96, 0.97) | 0.89 (0.85, 0.92) | 0.90 (0.86, 0.93) | 0.97 (0.96, 0.97) | 0.87 (0.84, 0.90) | 0.91 (0.88, 0.94) |
| | 100% + CheXpert | 0.96 (0.95, 0.97) | 0.88 (0.85, 0.91) | 0.91 (0.87, 0.93) | NA | NA | NA |
| Fracture | 20% | 0.91 (0.89, 0.93) | 0.88 (0.81, 0.94) | 0.78 (0.67, 0.84) | 0.86 (0.82, 0.90) | 0.71 (0.64, 0.79) | 0.88 (0.75, 0.93) |
| | 40% | 0.93 (0.90, 0.91) | 0.82 (0.75, 0.88) | 0.88 (0.80, 0.94) | 0.91 (0.88, 0.94) | 0.86 (0.78, 0.92) | 0.82 (0.67, 0.89) |
| | 60% | 0.94 (0.92, 0.96) | 0.89 (0.83, 0.94) | 0.88 (0.71, 0.91) | 0.94 (0.92, 0.96) | 0.83 (0.77, 0.89) | 0.90 (0.82, 0.95) |
| | 80% | 0.94 (0.91, 0.96) | 0.78 (0.71, 0.85) | 0.94 (0.86, 0.96) | 0.94 (0.92, 0.97) | 0.83 (0.76, 0.89) | 0.93 (0.78, 0.96) |
| | 100% | 0.93 (0.91, 0.96) | 0.88 (0.81, 0.94) | 0.88 (0.73, 0.90) | 0.96 (0.94, 0.98) | 0.91 (0.85, 0.95) | 0.89 (0.82, 0.93) |
| | 100% + CheXpert | 0.92 (0.90, 0.95) | 0.86 (0.78 0.92) | 0.83 (0.75, 0.89) | NA | NA | NA |
| Mass | 20% | 0.79 (0.76, 0.82) | 0.85 (0.79, 0.90) | 0.60 (0.51, 0.66) | 0.83 (0.80, 0.86) | 0.62 (0.55, 0.68) | 0.88 (0.80, 0.92) |
| | 40% | 0.86 (0.84, 0.89) | 0.68 (0.62, 0.74) | 0.89 (0.82, 0.92) | 0.88 (0.86, 0.91) | 0.72 (0.66, 0.78) | 0.89 (0.83, 0.93) |
| | 60% | 0.92 (0.90, 0.94) | 0.89 (0.84, 0.93) | 0.80 (0.73, 0.85) | 0.91 (0.89, 0.93) | 0.85 (0.80, 0.90) | 0.83 (0.72, 0.88) |
| | 80% | 0.92 (0.90, 0.94) | 0.81 (0.76, 0.86) | 0.88 (0.79, 0.91) | 0.92 (0.90, 0.94) | 0.85 (0.80, 0.90) | 0.84 (0.76, 0.89) |
| | 100% | 0.92 (0.90, 0.94) | 0.82 (0.77, 0.87) | 0.86 (0.80, 0.90) | 0.93 (0.91, 0.95) | 0.85 (0.79, 0.89) | 0.87 (0.81, 0.90) |
| | 100% + CheXpert | 0.89 (0.87, 0.91) | 0.76 (0.70, 0.82) | 0.85 (0.78, 0.90) | NA | NA | NA |
| Nodule | 20% | 0.76 (0.73, 0.79) | 0.84 (0.79, 0.88) | 0.55 (0.46, 0.59) | 0.77 (0.74, 0.80) | 0.62 (0.56, 0.67) | 0.82 (0.73, 0.85) |
| | 40% | 0.84 (0.82, 0.86) | 0.76 (0.72, 0.81) | 0.76 (0.70, 0.80) | 0.83 (0.80, 0.85) | 0.80 (0.75, 0.84) | 0.74 (0.64, 0.78) |
| | 60% | 0.84 (0.82, 0.86) | 0.78 (0.72, 0.82) | 0.76 (0.71, 0.79) | 0.85 (0.83, 0.87) | 0.74 (0.69, 0.79) | 0.79 (0.73, 0.84) |
| | 80% | 0.86 (0.84, 0.88) | 0.69 (0.64, 0.74) | 0.85 (0.79, 0.89) | 0.86 (0.84, 0.89) | 0.76 (0.72, 0.81) | 0.82 (0.76, 0.86) |
| | 100% | 0.87 (0.85, 0.89) | 0.81 (0.76, 0.85) | 0.76 (0.70, 0.80) | 0.86 (0.84, 0.88) | 0.74 (0.69, 0.79) | 0.82 (0.77, 0.86) |
| | 100% + CheXpert | 0.79 (0.77, 0.82) | 0.73 (0.68, 0.78) | 0.73 (0.66, 0.77) | NA | NA | NA |
| Pneumonia | 20% | 0.85 (0.83, 0.87) | 0.73 (0.69, 0.78) | 0.79 (0.74, 0.83) | 0.82 (0.80, 0.84) | 0.79 (0.74, 0.83) | 0.70 (0.65, 0.73) |
| | 40% | 0.86 (0.84, 0.88) | 0.78 (0.73, 0.82) | 0.79 (0.74, 0.83) | 0.86 (0.84, 0.88) | 0.78 (0.73, 0.82) | 0.78 (0.74, 0.82) |
| | 60% | 0.87 (0.845, 0.88) | 0.78 (0.74, 0.82) | 0.79 (0.74 0.83) | 0.87 (0.86, 0.89) | 0.80 (0.76, 0.84) | 0.77 (0.72, 0.80) |
| | 80% | 0.88 (0.86, 0.90) | 0.80 (0.76, 0.84) | 0.80 (0.73, 0.83) | 0.89 (0.87, 0.90) | 0.73 (0.69, 0.78) | 0.87 (0.82, 0.90) |
| | 100% | 0.89 (0.87, 0.90) | 0.83 (0.79, 0.87) | 0.81 (0.76, 0.84) | 0.89 (0.87, 0.90) | 0.84 (0.79, 0.87) | 0.78 (0.71, 0.80) |
| | 100% + CheXpert | 0.88 (0.86, 0.90) | 0.79 (0.74, 0.83) | 0.80 (0.75, 0.84) | NA | NA | NA |
| Pneumothorax | 20% | 0.92 (0.90, 0.94) | 0.81 (0.76, 0.86) | 0.90 (0.84, 0.93) | 0.95 (0.93, 0.96) | 0.86 (0.81, 0.90) | 0.93 (0.88, 0.96) |
| | 40% | 0.96 (0.94, 0.97) | 0.89 (0.84, 0.93) | 0.89 (0.83, 0.92) | 0.96 (0.94, 0.97) | 0.89 (0.84, 0.93) | 0.93 (0.87, 0.96) |
| | 60% | 0.96 (0.94, 0.97) | 0.88 (0.81, 0.92) | 0.91 (0.87, 0.95) | 0.96 (0.95, 0.98) | 0.90 (0.86, 0.94) | 0.91 (0.80, 0.94) |
| | 80% | 0.97 (0.96, 0.98) | 0.92 (0.88, 0.96) | 0.94 (0.88, 0.95) | 0.97 (0.96, 0.98) | 0.92 (0.87, 0.96) | 0.94 (0.90, 0.96) |
| | 100% | 0.97 (0.96, 0.98) | 0.91 (0.87, 0.95) | 0.91 (0.84, 0.94) | 0.97 (0.95, 0.98) | 0.92 (0.88, 0.95) | 0.93 (0.84, 0.96) |
| | 100% + CheXpert | 0.96 (0.95, 0.98) | 0.92 (0.88, 0.96) | 0.87 (0.83, 0.91) | NA | NA | NA |
| Tuberculosis | 20% | 0.91 (0.90, 0.93) | 0.80 (0.76, 0.83) | 0.89 (0.85, 0.91) | 0.90 (0.89, 0.91) | 0.82 (0.79, 0.85) | 0.85 (0.81, 0.87) |
| | 40% | 0.93 (0.92, 0.94) | 0.87 (0.83, 0.90) | 0.85 (0.82 0.88) | 0.92 (0.91, 0.93) | 0.87 (0.83, 0.90) | 0.83 (0.80, 0.86) |
| | 60% | 0.93 (0.92, 0.94) | 0.81 (0.78, 0.85) | 0.89 (0.86, 0.91) | 0.93 (0.92, 0.94) | 0.89 (0.86, 0.92) | 0.82 (0.78, 0.85) |
| | 80% | 0.94 (0.93, 0.95) | 0.90 (0.87, 0.92) | 0.84 (0.80, 0.87) | 0.93 (0.92, 0.94) | 0.87 (0.84, 0.90) | 0.86 (0.83, 0.88) |
| | 100% | 0.94 (0.93, 0.95) | 0.86 (0.83, 0.88) | 0.88 (0.85, 0.91) | 0.94 (0.93, 0.95) | 0.87 (0.84, 0.90) | 0.86 (0.83, 0.89) |
| | 100% + CheXpert | 0.93 (0.92, 0.94) | 0.85 (0.82, 0.88) | 0.86 (0.81, 0.89) | NA | NA | NA |
| Aortic Calcification | 20% | 0.85 (0.83, 0.88) | 0.91 (0.86, 0.95) | 0.66 (0.49, 0.70) | 0.94 (0.91, 0.96) | 0.88 (0.83, 0.93) | 0.89 (0.74, 0.93) |
| | 40% | 0.96 (0.94, 0.97) | 0.85 (0.79, 0.90) | 0.95 (0.88, 0.97) | 0.96 (0.95, 0.98) | 0.93 (0.87, 0.97) | 0.93 (0.74, 0.94) |
| | 60% | 0.97 (0.96, 0.98) | 0.89 (0.84, 0.93) | 0.95 (0.89, 0.97) | 0.96 (0.95, 0.98) | 0.92 (0.87, 0.96) | 0.94 (0.76, 0.96) |
| | 80% | 0.98 (0.97, 0.99) | 0.97 (0.90, 0.97) | 0.91 (0.87, 0.94) | 0.98 (0.97, 0.99) | 0.94 (0.90, 0.97) | 0.93 (0.82, 0.96) |
| | 100% | 0.98 (0.98, 0.99) | 0.93 (0.89, 0.96) | 0.96 (0.88, 0.97) | 0.97 (0.96, 0.99) | 0.95 (0.91, 0.98) | 0.92 (0.74, 0.95) |
| | 100% + CheXpert | 0.96 (0.95, 0.97) | 0.93 (0.88, 0.96) | 0.86 (0.77, 0.89) | NA | NA | NA |

A detailed descriptions of the model performance in Figure 4. The cutting points for sensitivities and specificities are computed from tuning set. AUC: Area under receiver receiver operating characteristic curve. CI: confidence interval.

**Table E2. Detailed statistics of JAFROC-FOMs of CheXNet and CheXDet on DS1.**

| Disease | CheXNet$_{100}$ JAFROC-FOM (95% CI) | The ratio of training data for CheXDet | CheXDet JAFROC-FOM (95% CI) |
|---|---|---|---|
| Aortic Calcification | 0.083 (0.076, 0.090) | 20% | 0.86 (0.82, 0.89) |
| | | 40% | 0.92 (0.89, 0.94) |
| | | 60% | 0.91 (0.89, 0.94) |
| | | 80% | 0.94 (0.91, 0.96) |
| | | 100% | *0.95 (0.93, 0.97) |
| Cardiomegaly | 0.280 (0.256, 0.303) | 20% | 0.92 (0.90, 0.94) |
| | | 40% | 0.94 (0.92, 0.96) |
| | | 60% | 0.95 (0.93, 0.97) |
| | | 80% | 0.96 (0.94, 0.97) |
| | | 100% | *0.96 (0.95, 0.98) |
| Pleural Effusion | 0.269 (0.249, 0.289) | 20% | 0.64 (0.62, 0.66) |
| | | 40% | 0.68 (0.65, 0.70) |
| | | 60% | 0.71 (0.69, 0.73) |
| | | 80% | 0.73 (0.71, 0.75) |
| | | 100% | *0.73 (0.71, 0.75) |
| Fracture | 0.347 (0.339, 0.355) | 20% | 0.56 (0.52, 0.59) |
| | | 40% | 0.62 (0.58, 0.66) |
| | | 60% | 0.64 (0.60, 0.69) |
| | | 80% | 0.68 (0.63, 0.72) |
| | | 100% | *0.72 (0.67, 0.76) |
| Mass | 0.483 (0.473, 0.492) | 20% | 0.67 (0.63, 0.70) |
| | | 40% | 0.74 (0.71, 0.78) |
| | | 60% | 0.76 (0.73, 0.80) |
| | | 80% | 0.78 (0.74, 0.81) |
| | | 100% | *0.81 (0.78, 0.84) |
| Nodule | 0.404 (0.397, 0.411) | 20% | 0.53 (0.50, 0.56) |
| | | 40% | 0.55 (0.51, 0.58) |
| | | 60% | 0.59 (0.55, 0.62) |
| | | 80% | 0.61 (0.57, 0.64) |
| | | 100% | *0.63 (0.59, 0.66) |
| Pneumonia | 0.134 (0.121, 0.147) | 20% | 0.53 (0.50, 0.56) |
| | | 40% | 0.56 (0.53, 0.59) |
| | | 60% | 0.56 (0.53, 0.59) |
| | | 80% | 0.60 (0.57, 0.63) |
| | | 100% | *0.59 (0.57, 0.62) |
| Pneumothorax | 0.134 (0.104, 0.163) | 20% | 0.78 (0.74, 0.81) |
| | | 40% | 0.82 (0.79, 0.85) |
| | | 60% | 0.84 (0.81, 0.87) |
| | | 80% | 0.86 (0.84, 0.89) |
| | | 100% | *0.87 (0.84, 0.90) |
| Tuberculosis | 0.462 (0.457, 0.467) | 20% | 0.64 (0.61, 0.66) |
| | | 40% | 0.66 (0.64, 0.68) |
| | | 60% | 0.69 (0.67, 0.71) |
| | | 80% | 0.70 (0.68, 0.72) |
| | | 100% | *0.73 (0.71, 0.75) |

A detailed descriptions of the model performance in Figure 6. Best performance is highlighted with *. JAFROC-FOM: free-response receiver-operating characteristic figure of merit. CI: confidence interval.